\documentclass[RNAAS]{aastex63}

\graphicspath{{./}{figures/}}

\newcommand{\adra}{\mbox{$\alpha$~Dra}}
\newcommand{\kms}{\mbox{km\,s$^{-1}$}}
\newcommand{\vsini}{\mbox{$v\sin i$}}

\newcommand{\lboo}{\mbox{$\lambda$~Bootis}}

\newcommand{\tess}{{\em TESS\/}}

\begin{document}

\title{A dance with dragons: \textit{TESS} reveals $\alpha$~Draconis is a detached eclipsing binary}

\correspondingauthor{Timothy R. Bedding}
\email{tim.bedding@sydney.edu.au}

\author[0000-0001-5222-4661]{Timothy R. Bedding}
\author[0000-0003-3244-5357]{Daniel R. Hey}
\author[0000-0002-5648-3107]{Simon J. Murphy}


\affiliation{Sydney Institute for Astronomy, School of Physics, University of Sydney 2006, Australia}
\affiliation{Stellar Astrophysics Centre, Aarhus University, DK-8000 Aarhus C, Denmark}

\keywords{Stellar photometry --- Eclipsing binary stars}

\section{} 

Detached eclipsing binaries allow stellar masses and radii to be measured with unrivalled accuracy \citep{Andersen1991,Torres++2010}.  While inspecting light curves obtained with the \textit{Transiting Exoplanet Survey Satellite} (\tess; \citealt{Ricker++2015}), we noticed that the A0\,III star \adra\ shows clear and well-separated primary and secondary eclipses.  With a $V$ magnitude of $3.68$, \adra\ is brighter than most previously known detached eclipsing binaries \citep{Malkov++2006,Zasche++2009,Avvakumova++2013}, with the only brighter systems apparently being $\gamma$~Per \citep{Griffin2007} and $\delta$~Vel~A \citep{Pribulla++2011}.

\begin{figure*}
\centering
\includegraphics[width=\linewidth]{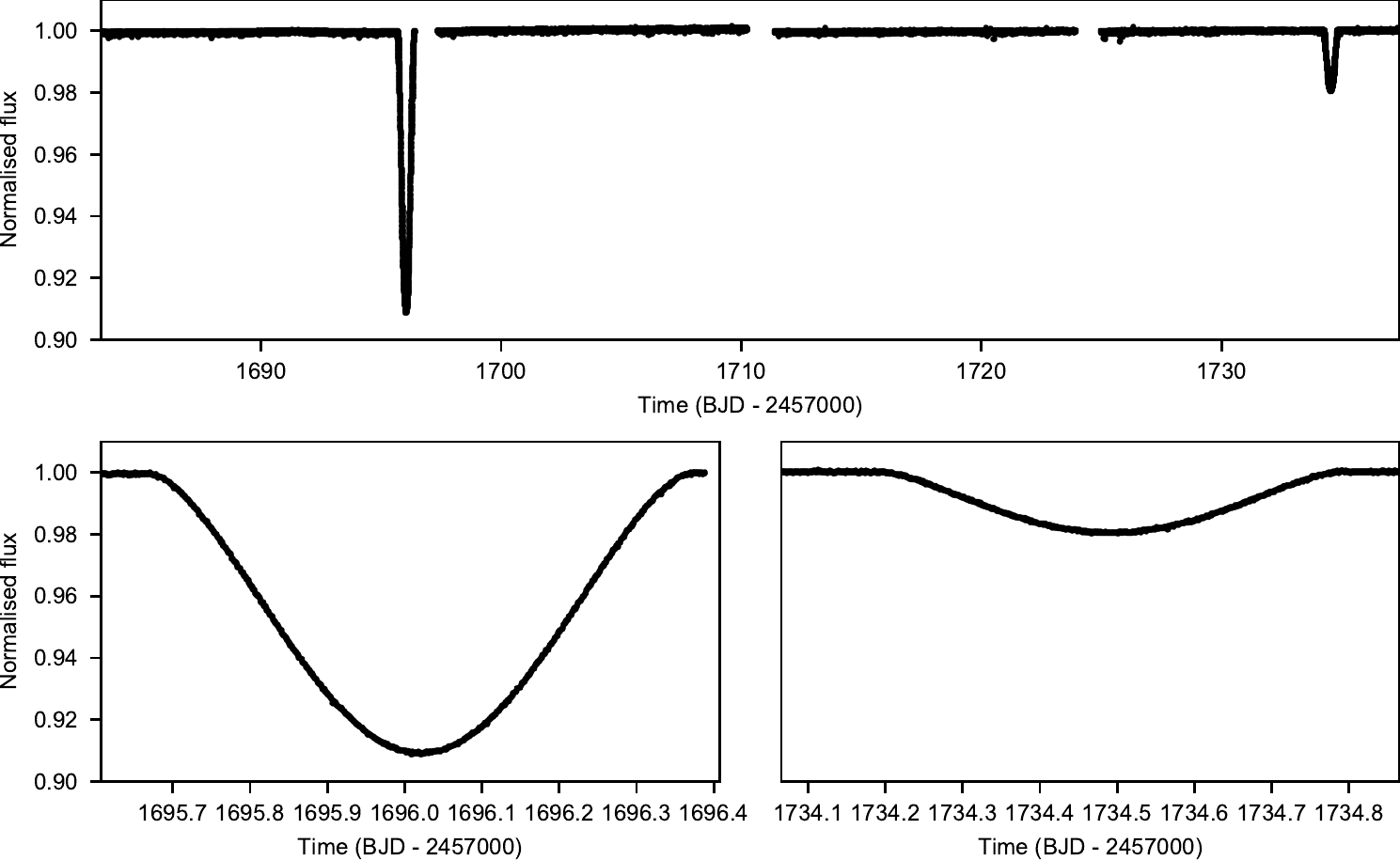}
\caption{\tess\ light curve of \adra\ at 2-minute cadence using the Simple Aperture Photometry (SAP) flux, show the primary and secondary eclipses. The bottom panels show close-up views (both 0.8\,d wide) of the primary and secondary eclipses (left and right, respectively).}
    \label{fig:adra}
\end{figure*}

The star \adra\ (Thuban; HR~5291; HD~123299; HIP~68756) is a well-studied single-lined spectroscopic binary, with a period of 51.5\,d and an eccentricity of 0.43 \citep[e.g.,][]{Harper1907,Elst+Nelles1983,Budovicov++2004,Bischoff++2017}. 
The currently available \tess\ observations for \adra\ cover two 27-d sectors\footnote{\url{https://mast.stsci.edu/portal/Mashup/Clients/Mast/Portal.html}}.  The light curve (Fig.~\ref{fig:adra}) shows a primary eclipse in Sector 14 and a secondary eclipse in Sector~15, separated in time by 38.5\,d.
Optical interferometry shows that the angular separation is a few milliarcsec and the magnitude difference at 700\,nm is $1.83 \pm 0.07$ \citep{Hutter++2016}.  This flux ratio implies that a total eclipse would have a depth of 16\%.  The two \tess\ eclipses have depths of 9\%\ and 2\%, indicating that the eclipses are partial and the inclination is slightly less than 90$^\circ$.

The star lies close to, but not inside, the \tess\ continuous viewing zone (CVZ) and further observations will be available from Sectors 16, 21 and 22\footnote{\url{https://heasarc.gsfc.nasa.gov/cgi-bin/tess/webtess/wtv.py}}.  
We expect the next primary eclipse to occur at BJD 2458747.4, which falls within Sector~16.  Another secondary eclipse will fall in Sector~21 and another primary eclipse will fall in Sector~22.

Spectroscopically, \adra\ is used as the MK standard for spectral type A0~III \citep{Gray++1987,Gray+Corbally2009}.  
Although \adra\ has been reported as a \lboo\ star (showing surface chemical abundances that imply accretion from circumstellar material), it is not a member of this class \citep{Murphy++2015}.
Its projected rotation velocity is quite low for this spectral type ($\vsini \sim 26\,\kms$; \citealt{Gray2014}).  Now that we know the system is eclipsing, the assumption of reasonable alignment between the rotation and orbital axes implies that the primary is indeed a slow rotator (rather than being a rapid rotator seen pole-on, like the A0 V star, Vega; \citealt{Hill+2010}).


\citet{Kallinger++2004} suggested that \adra\ is photometrically variable, with a period of about 53\,min and an amplitude of 1--2\,mmag.  They speculated that \adra\ could belong to the unconfirmed class of so-called Maia variables \citep[for recent discussions, see][]{Balona++2015,Szewczuk++2017,White++2017}.  It was this possibility led us to examine the \tess\ light curve for this star.  However, apart from the eclipses, the light curve shows no evidence for variability and we can rule out variability on timescales shorter than 8\,hr at the level of 10\,ppm (parts per million).

\acknowledgments

We thank the \tess\ team for making this research possible, and  Daniel Huber for helpful comments.  We gratefully acknowledge support from the Australian Research Council, and from the Danish National Research Foundation (Grant DNRF106) through its funding for the Stellar Astrophysics Centre (SAC). DRH acknowledges the support of the Australian Government Research Training Program (AGRTP).

\facility{TESS}



\end{document}